\documentclass[twocolumn,superscriptaddress,prb]{revtex4-2}

\usepackage{dcolumn}
\usepackage{bm}
\usepackage{subfigure}
\usepackage[normalem]{ulem}
\usepackage{enumerate}
\usepackage{cancel}
\usepackage{flushend}
\usepackage{bbold}
\usepackage{mathtools}
\usepackage{float}
\usepackage{stmaryrd}
\usepackage{colortbl}
\usepackage[table]{xcolor}
\usepackage{array,ragged2e}
\usepackage{amssymb}
\usepackage{wasysym,graphicx,multirow,textcomp}
\usepackage{url}
\usepackage{romannum}
\usepackage{tabularx}

\usepackage{changes}

\usepackage[colorlinks = true,linkcolor = red,citecolor = magenta]{hyperref}
\usepackage[sort&compress]{natbib}

\usepackage{tikz-cd}

\newcommand{\mh}{\mathcal{H}}
\newcommand{\sgg}{\mathcal{G}}
\newcommand{\efb}{E_\mathrm{FB}}

\makeatletter
\newcommand{\thickhline}{%
    \noalign {\ifnum 0=`}\fi \hrule height 1pt
    \futurelet \reserved@a \@xhline
}
\newcolumntype{"}{@{\hskip\tabcolsep\vrule width 1pt\hskip\tabcolsep}}
\makeatother

\newcommand{\pcsadd}{Center for Theoretical Physics of Complex Systems, Institute for Basic Science (IBS), Daejeon 34126, Republic of Korea}
\newcommand{\ustadd}{Basic Science Program, Korea University of Science and Technology (UST), Daejeon 34113, Republic of Korea}

\begin{document}

\draft

\title{Symmetry-protected flatband conditions for Hamiltonians with local symmetry}

\author{Jung-Wan Ryu}
    \address{\pcsadd}
    \address{\ustadd}

\author{Alexei Andreanov}
    \email{aalexei@ibs.re.kr}
    \address{\pcsadd}
    \address{\ustadd}

\author{Hee Chul Park}
    \email{hc2725@gmail.com}
    \address{Department of Physics, Pukyong National University, Busan 48513, Republic of Korea}

\author{Jae-Ho Han}
    \email{jaehohan@ibs.re.kr}
    \address{\pcsadd}

\date{\today}

\begin{abstract}
    We derive symmetry-based conditions for tight-binding Hamiltonians with flatbands to have compact localized eigenstates occupying a single unit cell. 
    The conditions are based on unitary operators commuting with the Hamiltonian and are associated with local symmetries that guarantee compact localized states and a flatband. 
    We illustrate the conditions for compact localized states and flatbands with simple Hamiltonians with given symmetries. 
    We also apply the results to general cases such as a Hamiltonian with long-range hopping and a higher-dimensional Hamiltonian.
\end{abstract}

\maketitle

\section{Introduction}
\label{sec:introduction}

Hamiltonians with discrete translational symmetry can have dispersionless bands known as \emph{flatbands}, with energy spectra \(E(\bm k)\) independent of momentum \(\bm k\)~\cite{Ber13, Par13, Fla14, Der15, Ley18}. 
The flatness of a band implies zero group velocity, infinite effective mass of electrons, and suppressed electron and wave transport. 
The origin of flatbands is destructive interference due to a fine-tuning of the hopping or lattice symmetry.
An important property of flatbands in short-range Hamiltonians is \emph{compact localized states} (CLSs)---flatband eigenstates that are perfectly localized on a finite number of lattice sites.
This is in contrast to Anderson localization where eigenstates are localized exponentially over the entire lattice~\cite{And58}. 
Since the first report of a flatband in a dice lattice~\cite{Sut86}, a variety of flatband models have been identified, e.g., Lieb~\cite{Lie89, She10, Wee10, Apa10, Gol11, Asa11}, kagome~\cite{Tho82, Hal88, Kan05, Guo09, Xu15, Ye18, Ye19}, and honeycomb~\cite{Wu07, Cas09, Kal14, Jac14} lattices. 
Despite their fine-tuned character and strong sensitivity to perturbations, flatbands have been realized in multiple experiments in different settings: superconducting networks~\cite{Vid98, Abi99}, photonic flatbands~\cite{Sza06, Guz14, Vic15, Muk15, Ley18-2}, optical lattices for cold atoms~\cite{Tai15, Oza17, Tai20, Leu20, Kan20}, and engineered atomic lattices~\cite{Dro17, Slo17, Hud20}.

Symmetry, one of the fundamental principles of physics, allows us to predict certain properties of a system without solving the often complicated underlying equations.
In quantum mechanics, symmetry is associated with an operator that commutes with a Hamiltonian. 
A symmetry is \emph{global} if the respective operator is \emph{independent} of lattice sites, and a symmetry is \emph{local} if the operator is \emph{dependent} on lattice sites.
It has been discovered that certain classes of local symmetries can indeed be systematically linked to CLSs and flatbands~\cite{Fla14, Ram17, Roe18}. 
Although it is well known that flatbands and CLSs result from destructive interference caused by fine-tuning~\cite{Tho71} or by  specific symmetries~\cite{Sut86, Lie89}, a relation between fine-tuning and symmetry has not been fully established yet~\cite{cualuguaru2022general,bae2023isolated}. 
In this work, we derive the exact conditions for CLSs occupying a single unit cell and the corresponding flatbands from the global and local symmetries of the system.
We then propose a method to design lattice Hamiltonians with flatbands in terms of given symmetries and corresponding unitary operators. 
As a result, we demonstrate that if a Hamiltonian possesses a local symmetry for which the associated unitary operators are also operators of a global symmetry of the Hamiltonian, such Hamiltonian should have at least one compact localized state and corresponding flatband.

This paper is structured as follows.
In Section II, we derive the conditions for CLSs and corresponding flatbands in a Hamiltonian with discrete translational symmetry in terms of both global and local symmetries. 
In Section III, we illustrate the method with several simple examples, including already-known flatband models.
Section IV introduces generalizations of our method to longer-range hopping and higher dimensions. 
In Section V, we summarize and discuss our results.

\section{Flatbands generated by symmetries}
\label{sec:flatbands}

Consider a one-dimensional (1D) tight-binding model with nearest-neighbor unit cell hopping.
The Hamiltonian in the second quantized form is given by
\begin{align}
    \mh &=& \sum_{i=1}^N \sum_{\alpha,\beta = 1}^m \Big( \hat c_{i\alpha}^\dagger H_{0;\alpha \beta} \hat c_{i\beta} + \hat c_{i+1 \alpha}^\dagger H_{1;\alpha\beta} \hat c_{i\beta} \notag \\
    \label{eq:ham}
    & &~~~~~~~~~~~ + \hat c_{i\alpha}^\dagger H_{1;\beta\alpha}^* \hat c_{i+1 \beta} \Big) \\
    &=& \sum_i \Big( \hat c_i^\dagger H_0 \hat c_i + \hat c_{i+1}^\dagger H_1 \hat c_i + \hat c_i^\dagger H_1^\dagger \hat c_{i+1} \Big), \notag
\end{align}
where \(\alpha = 1, 2, \cdots, m\) labels the basis in the unit cell, \(\hat c_{i\alpha}\) (\(\hat c_{i\alpha}^\dagger\)) is an annihilation (creation) operator at unit cell \(i\) and basis site \(\alpha\), and \(N\) is the total number of cells.
\(H_{0;\alpha\beta}\) are on-site energies (\(\alpha=\beta\)) and hopping constants (\(\alpha\neq\beta\)) within the unit cell, and \(H_{1;\alpha\beta}\) are inter-cell hoppings. 
The matrix notation with respect to the basis \(\alpha\) is introduced in the second equation.
Since we are interested in single-particle models, the statistics of \(\hat c_{i\alpha}\) (fermionic or bosonic) is irrelevant.

Let us assume that a given matrix \(H_0\) has a nontrivial finite symmetry group \(\sgg\) of order larger than one, e.g., with more than one element. 
Then all matrices of a representation \(U_i\) of group \(\sgg\) commute with the intra-cell Hamiltonian
\begin{gather}
    [U_i(g), H_0] = 0, \ \ \ U_i(g)^\dagger U_i(g) = 1, \ \ \ \forall g \in \sgg.
    \label{eq:symm_unit}
\end{gather}
Here the subscript \(i\) indicates the lattice site.
Our goal is to systematically find \(H_1\) consistent with the symmetries \(\sgg\), such that the full Hamiltonian, Eq.~\eqref{eq:ham}, has flatbands.
This consistency condition has important implications when the group \(\sgg\) has more than one element.
The symmetry operation on the total system can be represented as a product of these operators, \(U_\mathrm{tot}(g) = \otimes_{i=1}^N U_i(g)\).
For a \emph{global} symmetry, the operator acting on site \(i\) \(U_i(g) = U(g)\) is the same for all sites.
Requesting that the Hamiltonian \(\mh\) is invariant under global symmetry, one obtains the condition
\begin{gather}
    [U(g), H_1] = 0,
    \label{Eq:symm_inter}
\end{gather}
in addition to the condition in Eq.~\eqref{eq:symm_unit}. 

\begin{figure}[tb]
    \centering
    \includegraphics[width=1.0\linewidth]{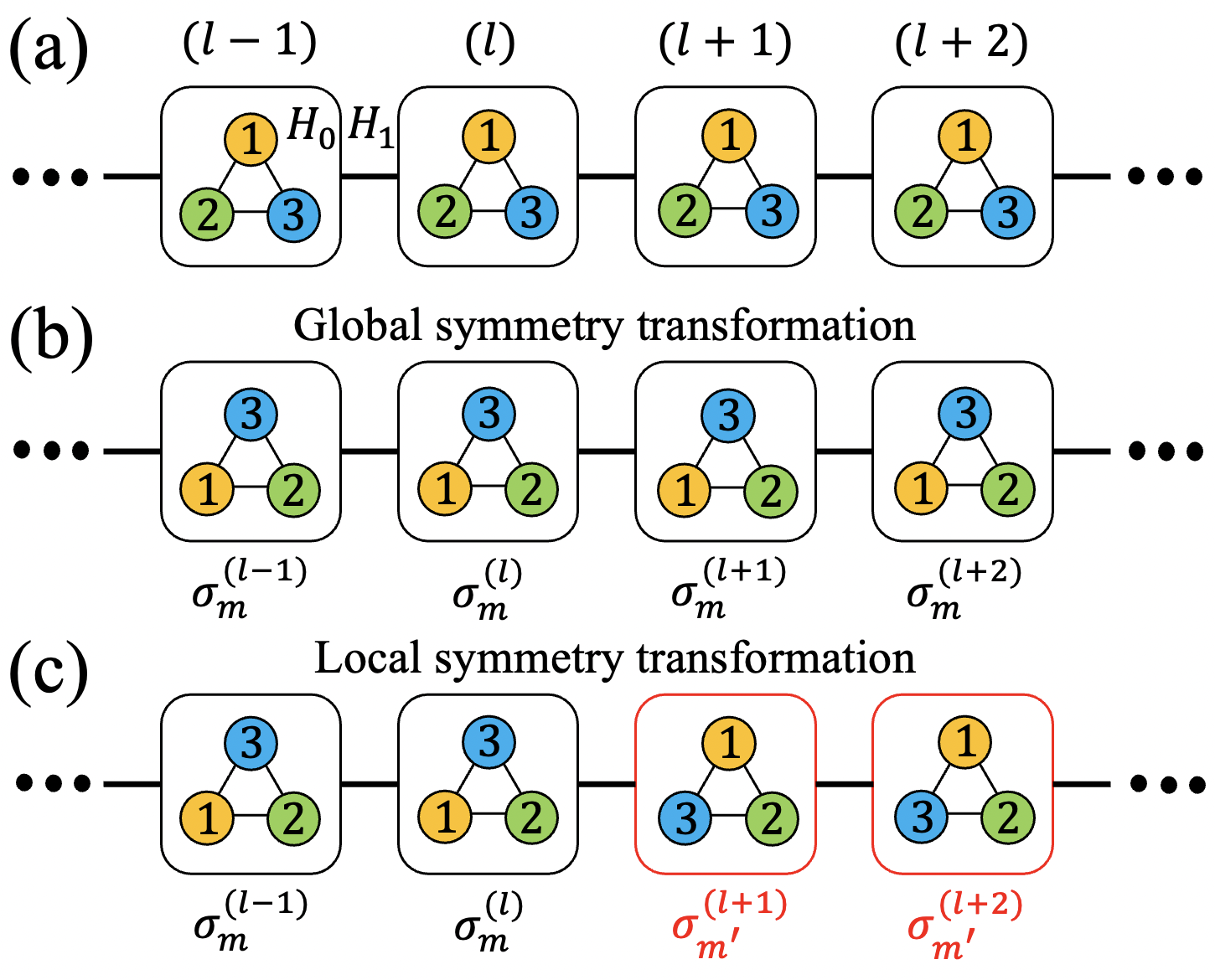}
    \caption{
        (a) Three-band lattice model with discrete translational symmetry. 
        \(H_0\) and \(H_1\) encode intra- and inter-cell hopping, respectively, and \((l)\) labels the unit cells.
        Schematic presentation of (b) \emph{global} and (c) \emph{local} symmetry transformations. 
        \(\sigma_m\) and \(\sigma_{m'}\) are unitary operators commuting with the total Hamiltonian, Eq.~\eqref{eq:ham}.
        For example, \(\sigma_{m}\) represents a \(120^\circ\) discrete rotation in the counterclockwise direction, and \(\sigma_{m'}\) is a reflection with respect to a vertical axis.
    }
    \label{fig:1}
\end{figure}

Now, consider two different symmetries of \(H_0\), \(g_1\) and \(g_2\), both elements of \(\sgg\).
Requesting that the total Hamiltonian is invariant under \(g_1\) and \(g_2\) globally, we have the following commutation relations,
\begin{gather}
    [\sigma_{1,2}, H_0] = 0, \ \ \ 
    [\sigma_{1,2}, H_1] = 0,
    \label{eq:comm}
\end{gather}
where \(\sigma_{1,2} = U(g_{1,2})\).
However, the presence of \emph{two} distinct elements allows to construct more complex symmetries.
In particular, to impose a CLS, we consider a \emph{kink} operation, namely a symmetry \(U_\mathrm{tot} = \otimes_{i=1}^N \sigma_i\) acting as \(\sigma_1\) for sites \(i \leq l\) and \(\sigma_2\) for \(i \geq l+1\), as shown in Fig.~\ref{fig:1}.
We now require that the total Hamiltonian is invariant under the kink operation.
Then at the location of the kink, we have the condition
\begin{gather}
    \sigma_2^\dagger H_1 \sigma_1 = H_1.
    \label{eq:CLS_condit}
\end{gather}
To show that this condition implies the existence of a CLS, we rewrite the condition as
\begin{gather}
    H_1 (1-\sigma_2^{-1} \sigma_1) = 0.
\end{gather}
Acting with this operator on eigenstates \(\phi_n\) of \(H_0\), we find
\begin{gather}
    H_1 \phi_n' = 0, \ \ \ \phi_n' \equiv (1-\sigma_2^{-1} \sigma_1) \phi_n.
\end{gather}
If \(\phi_n' = 0\) for all \(n\), then \((1-\sigma_2^{-1} \sigma_1) = 0\) or \(\sigma_1 = \sigma_2\), due to the completeness of the eigenstates \(\{\phi_n\}\).
This contradicts the assumption of two \emph{different} symmetries \(g_1\) and \(g_2\).
There is therefore at least one nontrivial state \(\phi_n'\) that is a zero mode of \(H_1\) and that is also an eigenstate of \(H_0\), since \(\sigma_{1,2}\) commute with \(H_0\) [Eq.~\eqref{eq:comm}].
Accordingly, the state \(\phi_n'\) is an eigenstate of the total Hamiltonian with \(H_1 \phi_n' = 0\), and so it is a CLS.

We note that the operator \(\sigma_i\) is determined up to phase; see Eq.~\eqref{eq:symm_unit}. 
The number of allowed phases is finite for a finite symmetry group.
Consider an element \(g\) of order \(N_g\),
\begin{gather}
    g^{N_g} = e, \ \ \ N_g \in \mathbb{N},
\end{gather}
where \(e\) is the identity element. 
Let \(U(g)\) be the operator corresponding to \(g\). 
Then the operator \(e^{i\theta} U(g)\), \(\theta \in \mathbb R\) also satisfies Eq.~\eqref{eq:symm_unit}.
However, the condition \(g^{N_g} = e\) restricts the values of \(\theta\), \(e^{iN_g\theta} = 1\), or \(\theta_n = 2\pi\frac{n}{N_g}\), \(n=0, 1, \cdots, N_g-1\).
Thus there can be \(N\) operators
\begin{gather}
    U(g), \ e^{i\theta_1} U(g), \ e^{i\theta_2} U(g), \cdots, e^{i\theta_{N_g-1}} U(g)
\end{gather}
corresponding to the element \(g\).
Due to the unitarity and homeomorphic property of \(U(g)\), the phase \(e^{i\theta}\) is the same as the eigenvalues of \(\sigma_g \equiv U(g)\).
Therefore, we can rewrite the flatband condition in Eq.~\eqref{eq:CLS_condit} as
\begin{gather}
    \lambda_2^i \sigma_1 H_1 = \lambda_1^j \sigma_2 H_1,
    \label{eq:final_sol5}
\end{gather}
where \(\lambda_{g}^{i}\) are eigenvalues of \(\sigma_{g}\) with \(i = 1, 2, 3,\cdots, N_g\).
This is the condition that is mainly used in the following sections.
From Eqs.~\eqref{eq:comm} and~\eqref{eq:final_sol5}, one can obtain the total Hamiltonian \(\mathcal H\) with flatbands.

\section{Obtaining \(H_1\) from given \(H_0\)}
\label{sec:examples}

We illustrate the generic method outlined above by considering a given intra-cell hopping matrix \(H_0\). 
First, we determine the symmetry group \(\sgg\) of \(H_0\), e.g., the unitary operators commuting with \(H_0\).
Then we identify inter-cell hopping matrices \(H_1\) from the given \(H_0\) and the associated unitary operators using the flatband condition derived in the previous section, Eq.~\eqref{eq:final_sol5}.
We note that setting \(H_1=0\) (or equivalent) is also a solution; however, this corresponds to the trivial case of disconnected unit cells and we do not consider such cases in the derivations below.

\subsection{\(2 \times 2\) Hamiltonian}
\label{subsec:examples1}

We start with the simplest possible setting in which flatbands can appear: a \(2 \times 2\) Hamiltonian with two bands,
\begin{gather}
    \label{eq:ham_1d}
    \mh = H_0 + e^{-i k} H_1 + e^{i k} H_1^{\dagger},
\end{gather}
with \(H_0\) having parity symmetry
\begin{gather}
    H_0 =  
    \begin{pmatrix}
        0 & 1 \\
        1 & 0 
    \end{pmatrix},
    H_1 =  
    \begin{pmatrix}
        t_{11} & t_{12} \\
        t_{21} & t_{22} 
    \end{pmatrix}.  
\end{gather}
Our purpose is to find \(H_1\) that satisfies both the global and local symmetry constraints of the given \(H_0\), and therefore $\mathcal H$ has at least one dispersionless energy band.
First, we identify the symmetries of \(H_0\): there are two unitary operators,
\begin{gather}
    \sigma_0 = 
    \begin{pmatrix}
        1 & 0 \\
        0 & 1
    \end{pmatrix}, \ 
    \sigma_1 = 
    \begin{pmatrix}
        0 & 1 \\
        1 & 0
    \end{pmatrix}, \
\end{gather}
commuting with \(H_0\).
We note that an identity matrix \(\sigma_0\) is always a unitary operator commuting with any Hamiltonian.

Given these symmetry generators, the flatband conditions for \(H_1\)[Eqs.~\eqref{eq:comm} and~\eqref{eq:final_sol5}] are given by
\begin{gather}
    \left[\sigma_0, H_1\right] = 0,
    \left[\sigma_1, H_1\right] = 0,
    \sigma_1 H_1 = \pm \sigma_0 H_1 = \pm H_1.
\end{gather}
These conditions can also be obtained considering three symmetry operators, \(\sigma_0\), \(\sigma_1\), and \(-\sigma_1\) instead of the two symmetry operators with phases.
Resolving the above flatband constraints with respect to \(t_{ij}\), we find the following hopping matrix:
\begin{gather}
    H_1 = 
    \begin{pmatrix}
        t_{11} & \pm t_{11} \\
        \pm t_{11} & t_{11} 
    \end{pmatrix}.
    \label{triv_22}
\end{gather}
This corresponds to the well-known case of a cross-stitch lattice with a single flatband~\cite{Fla14} for \(t_{11} \neq 0\) and all elements of the same sign. 

The other symmetry groups for the two-band case are studied systematically in Appendix~\ref{app:generalizations1}.

\subsection{\(3 \times 3\) Hamiltonian}
\label{subsec:examples2}

As the next example, consider a $3\times3$ Hamiltonian with symmetric all-to-all hopping within the unit cell,
\begin{gather}
    \mh = H_0 + e^{-i k} H_1 + e^{i k} H_1^{\dagger},
\end{gather}
where we choose
\begin{gather}
    H_0 =
    \begin{pmatrix}
        0 & 1 & 1 \\
        1 & 0 & 1 \\
        1 & 1 & 0
    \end{pmatrix},
    H_1 =  
    \begin{pmatrix}
        t_{11} & t_{12} & t_{13} \\
        t_{21} & t_{22} & t_{23} \\
        t_{31} & t_{32} & t_{33}
    \end{pmatrix}.
\end{gather}
The symmetry group \(\sgg\) of \(H_0\) consists of three symmetry operators,
\begin{gather}
    \sigma_0 = 
    \begin{pmatrix}
        1 & 0 & 0 \\
        0 & 1 & 0 \\
        0 & 0 & 1
    \end{pmatrix},
    \sigma_1 =  
    \begin{pmatrix}
        1 & 0 & 0 \\
        0 & 0 & 1 \\
        0 & 1 & 0
    \end{pmatrix},
    \sigma_2 =  
    \begin{pmatrix}
        0 & 1 & 0 \\
        0 & 0 & 1 \\
        1 & 0 & 0
    \end{pmatrix}.
\end{gather}
The flatband condition from Eq.~\eqref{eq:final_sol5} in this case is more complicated compared to the case of the \(2 \times 2\) Hamiltonian because there are three distinct unitary operators in the symmetry group that commute with \(H_0\). 
Conditions can be formulated for four different combinations of unitary operators: (\(\sigma_0\), \(\sigma_1\)), (\(\sigma_0\), \(\sigma_2\)), (\(\sigma_1\), \(\sigma_2\)), and (\(\sigma_0\), \(\sigma_1\), \(\sigma_2\)). 
Consequently, \(H_0\) satisfies the following relations:
\begin{align}
    \left[\sigma_0, H_0\right] &= 0, 
    \left[\sigma_1, H_0\right] = 0
    ~~ \mathrm{for} ~~ (\sigma_0, \sigma_1) \\
    \left[\sigma_0, H_0\right] &= 0,
    \left[\sigma_2, H_0\right] = 0 
    ~~ \mathrm{for} ~~ (\sigma_0, \sigma_2) \\
    \left[\sigma_1, H_0\right] &= 0,
    \left[\sigma_2, H_0\right] = 0 
    ~~ \mathrm{for} ~~ (\sigma_1, \sigma_2) \\
    \left[\sigma_0, H_0\right] &= 0,
    \left[\sigma_1, H_0\right] = 0,
    \left[\sigma_2, H_0\right] = 0 
    ~ \mathrm{for} ~ (\sigma_0, \sigma_1, \sigma_2).
\end{align}
The flatband conditions for \(H_1\) are 
\begin{widetext}
\begin{align}
    \left[\sigma_0, H_1\right] &= 0,
    \left[\sigma_1, H_1\right] = 0,
    \sigma_0 H_1 = \pm \sigma_1 H_1
    ~~~~~ \mathrm{for} ~~ (\sigma_0, \sigma_1) \\
    \left[\sigma_0, H_1\right] &= 0,
    \left[\sigma_2, H_1\right] = 0,
    \sigma_0 H_1 = e^{-i 2 j \pi / 3} \sigma_2 H_1 ~(j=0,1,2)
    ~~~~~ \mathrm{for} ~~ (\sigma_0, \sigma_2) \\
    \left[\sigma_1, H_1\right] &= 0,
    \left[\sigma_2, H_1\right] = 0,
    \sigma_1 H_1 = \pm e^{-i 2 j \pi / 3} \sigma_2 H_1 ~(j=0,1,2)
    ~~~~~ \mathrm{for} ~~ (\sigma_1, \sigma_2) \\
    \left[\sigma_0, H_1\right] &= 0,
    \left[\sigma_1, H_1\right] = 0,
    \left[\sigma_2, H_1\right] = 0, \\
    \sigma_0 H_1 &= \pm \sigma_1 H_1,  \sigma_0 H_1 = e^{-i 2 j \pi / 3} \sigma_2 H_1, \sigma_1 H_1 = \pm e^{-i 2 j \pi / 3} \sigma_2 H_1 ~(j=0,1,2)
   ~~~~~ \mathrm{for} ~~ (\sigma_0, \sigma_1, \sigma_2).
\end{align}
\end{widetext}
Resolving the above conditions with respect to the hopping matrix \(H_1\), parameterized as follows,
\begin{align}
    H_1 &=  
    \begin{pmatrix}
        t_{11} & t_{12} & t_{13} \\
        t_{21} & t_{22} & t_{23} \\
        t_{31} & t_{32} & t_{33}
    \end{pmatrix},
\end{align}
we find the following solutions,
\begin{align}
    H_1 &=
    \begin{pmatrix}
        t_{11} & t_{12} & t_{12} \\
        t_{21} & t_{22} & t_{22} \\
        t_{21} & t_{22} & t_{22}
    \end{pmatrix}
    \mathrm{or} 
    \begin{pmatrix}
        0 & 0 & 0 \\
        0 & t_{22} & -t_{22} \\
        0 & -t_{22} & t_{22}
    \end{pmatrix},
    ~~~~~ \mathrm{for} ~~ (\sigma_0, \sigma_1) \\
    H_1 &= 
    \begin{pmatrix}
        t_{11} & t_{11} & t_{11} \\
        t_{11} & t_{11} & t_{11} \\
        t_{11} & t_{11} & t_{11}
    \end{pmatrix}
    \mathrm{or} 
    \begin{pmatrix}
        t_{11} & e^{\mp i 2 \pi /3} t_{11} & e^{\pm i 2 \pi /3} t_{11} \\
        e^{\pm i 2 \pi /3} t_{11} & t_{11} & e^{\mp i 2 \pi /3} t_{11} \\
        e^{\mp i 2 \pi /3} t_{11} & e^{\pm i 2 \pi /3} t_{11} & t_{11}
    \end{pmatrix}
    \label{eq:triv_33}
\end{align}
for the remaining cases. 
For the choice of two unitary operators (\(\sigma_0\), \(\sigma_1\)), with the hopping \(t_{12} = 1\) (or \(t_{21} = 1\)) and all other hoppings set to zero for the first \(H_1\) in the above, the Hamiltonian corresponds to a diamond chain with a vertical link that has a tunable flatband~\cite{Fla14}.
If we consider special cases in which all matrix elements of \(H_1\) are equal to each other in Eq.~\eqref{eq:triv_33}, there is always only one dispersive band (see Appendix~\ref{app:trivialh1}).

\section{Generalizations}
\label{sec:generalizations}

Similarly as in the previous section, one can extend the analysis to the case of an arbitrary number of bands.
Our approach can be extended to longer-range hopping and higher dimensions, as we demonstrate below.

\subsection{Hamiltonian with long-range hopping}
\label{subsec:generalizations3}

We consider a 1D Hamiltonian with long-range hopping,
\begin{gather}
    H = H_0 + \sum_{j=1}^{L}\left(e^{-i k j} H_j + e^{i k j} H_j^{\dagger}\right),
    \label{LR_Ham}
\end{gather}
where $L$ is the longest hopping range. In the example case of three bands, the hopping matrices can be parameterized as
\begin{gather}
    H_0 = 
    \begin{pmatrix}
        \epsilon_{11} & \epsilon_{12} & \epsilon_{13} \\
        \epsilon_{21} & \epsilon_{22} & \epsilon_{23} \\
        \epsilon_{31} & \epsilon_{32} & \epsilon_{33}
    \end{pmatrix},
    H_j =
    \begin{pmatrix}
        t_{j11} & t_{j12} & t_{j13} \\
        t_{j21} & t_{j22} & t_{j23} \\
        t_{j31} & t_{j32} & t_{j33}
    \end{pmatrix}.
\end{gather}
We impose the following two symmetries on \(H_0\), expressed as unitary operators,
\begin{gather}
    \sigma_0 =  
    \begin{pmatrix}
        1 & 0 & 0 \\
        0 & 1 & 0 \\
        0 & 0 & 1
    \end{pmatrix},
    \sigma_1 = 
    \begin{pmatrix}
        1 & 0 & 0 \\
        0 & 0 & 1 \\
        0 & 1 & 0
    \end{pmatrix}.
    \label{eq:symm_long}
\end{gather}
Then \(H_0\) satisfies the relation
\begin{gather}
    \left[\sigma_0, H_0\right] = \left[\sigma_1, H_0\right] = 0,
\end{gather}
and the flatband conditions for \(H_1\) read
\begin{gather}
    \left[\sigma_0, H_j\right] = 0,
    \left[\sigma_1, H_j\right] = 0,
    \sigma_1 H_j = \pm \sigma_0 H_j = \pm H_j.
\end{gather}
From these constraints, it follows that
\begin{align}
\label{eq:h0_longrange}
    H_0 &=
    \begin{pmatrix}
        \epsilon_{11} & \epsilon_{12} & \epsilon_{12} \\
        \epsilon_{21} & \epsilon_{22} & \epsilon_{23} \\
        \epsilon_{21} & \epsilon_{23} & \epsilon_{22}
    \end{pmatrix}
\end{align}
and
\begin{align}
\label{eq:h1_longrange}
    H_j &=
    \begin{pmatrix}
        t_{j11} & t_{j12} & t_{j12} \\
        t_{j21} & t_{j22} & t_{j22} \\
        t_{j21} & t_{j22} & t_{j22}
    \end{pmatrix}
    \mathrm{or}
    \begin{pmatrix}
        0 & 0 & 0 \\
        0 & t_{j22} & -t_{j22} \\
        0 & -t_{j22} & t_{j22}
    \end{pmatrix}.
\end{align}
There is one flatband with \(\efb = \epsilon_{22} - \epsilon_{23}\) for the first choice of \(H_j\), or two flatbands with \(\efb = (\epsilon_{11}+\epsilon_{22}+\epsilon_{23} \mp \sqrt{8 \epsilon_{12} \epsilon_{21} + (-\epsilon_{11}+\epsilon_{22}+\epsilon_{23})^2})/2\) for the second choice of \(H_j\). 
An example of a 1D flatband Hamiltonian with next-nearest hopping terms derived using our method is presented in Appendix~\ref{app:ham_1d_lr}.

\subsection{2D Hamiltonian}
\label{subsec:generalizations2}

A two-dimensional (2D) generalization of the 1D Hamiltonian in Eq.~\eqref{eq:ham_1d} is given by
\begin{gather}
    H = H_0 + e^{-i k_{x}} H_1 + e^{i k_{x}} H_1^{\dagger} + e^{-i k_{y}} H_2 + e^{i k_{y}} H_2^{\dagger}.
    \label{2D_Ham}
\end{gather}
In the example case of three bands, the hopping matrices can be parameterized as
\begin{gather}
    H_0 = 
    \begin{pmatrix}
        \epsilon_{11} & \epsilon_{12} & \epsilon_{13} \\
        \epsilon_{21} & \epsilon_{22} & \epsilon_{23} \\
        \epsilon_{31} & \epsilon_{32} & \epsilon_{33}
    \end{pmatrix}, \\
    H_1 =
    \begin{pmatrix}
        t_{11} & t_{12} & t_{13} \\
        t_{21} & t_{22} & t_{23} \\
        t_{31} & t_{32} & t_{33}
    \end{pmatrix},
    H_2 =
    \begin{pmatrix}
        s_{11} & s_{12} & s_{13} \\
        s_{21} & s_{22} & s_{23} \\
        s_{31} & s_{32} & s_{33}
    \end{pmatrix}.
    \notag
\end{gather}
We impose the following two symmetries on \(H_0\), expressed as unitary operators,
\begin{gather}
    \sigma_0 =  
    \begin{pmatrix}
        1 & 0 & 0 \\
        0 & 1 & 0 \\
        0 & 0 & 1
    \end{pmatrix},
    \sigma_1 = 
    \begin{pmatrix}
        1 & 0 & 0 \\
        0 & 0 & 1 \\
        0 & 1 & 0
    \end{pmatrix},
    \label{eq:symm_2D}
\end{gather}
and coinciding with the same symmetries imposed in the previous example.
Then \(H_0\) satisfies the relation
\begin{gather}
    \left[\sigma_0, H_0\right] = \left[\sigma_1, H_0\right] = 0
\end{gather}
and the flatband conditions [Eq.~\eqref{eq:final_sol5}] for \(H_1\) read
\begin{gather}
    \left[\sigma_0, H_1\right] = 0,
    \left[\sigma_1, H_1\right] = 0,
    \sigma_1 H_1 = \pm \sigma_0 H_1 = \pm H_1, \\
    \left[\sigma_0, H_2\right] = 0,
    \left[\sigma_1, H_2\right] = 0,
    \sigma_1 H_2 = \pm \sigma_0 H_2 = \pm H_2.
\end{gather}
From these constraints, \(H_0\) follows as
\begin{align}
\label{eq:h0_2d}
    H_0 &= 
    \begin{pmatrix}
        \epsilon_{11} & \epsilon_{12} & \epsilon_{12} \\
        \epsilon_{21} & \epsilon_{22} & \epsilon_{23} \\
        \epsilon_{21} & \epsilon_{23} & \epsilon_{22}
    \end{pmatrix}.
\end{align}
There are two distinct solutions for \(H_{1,2}\).
The first one gives
\begin{align}
\label{eq:h1_2d}
    H_1 =
    \begin{pmatrix}
        t_{11} & t_{12} & t_{12} \\
        t_{21} & t_{22} & t_{22} \\
        t_{21} & t_{22} & t_{22}
    \end{pmatrix},
     H_2 =
    \begin{pmatrix}
        s_{11} & s_{12} & s_{12} \\
        s_{21} & s_{22} & s_{22} \\
        s_{21} & s_{22} & s_{22}
    \end{pmatrix},
\end{align}
having one flatband with \(\efb = \epsilon_{22} - \epsilon_{23}\).
The second solution for \(H_{1,2}\) reads
\begin{align}
\label{eq:h1_2d_2}
    H_1 =
    \begin{pmatrix}
        0 & 0 & 0 \\
        0 & t_{22} & -t_{22} \\
        0 & -t_{22} & t_{22}
    \end{pmatrix},
    H_2 =
    \begin{pmatrix}
        0 & 0 & 0 \\
        0 & s_{22} & -s_{22} \\
        0 & -s_{22} & s_{22}
    \end{pmatrix},
\end{align}
having two flatbands with \(\efb = (\epsilon_{11}+\epsilon_{22}+\epsilon_{23} \mp \sqrt{8 \epsilon_{12} \epsilon_{21} + (-\epsilon_{11}+\epsilon_{22}+\epsilon_{23})^2})/2\).
We note that these flatbands are the same as those in the previous 1D long-range Hamiltonian case.

An example of a combined 2D Hamiltonian with a cross-stitch chain along the x-axis and a tunable diamond chain along the y-axis is presented in Appendix~\ref{app:combined}.

\section{Summary}
\label{sec:summary}

We derived the conditions for lattice Hamiltonians with flatbands to have compact localized eigenstates that localize perfectly in a single unit cell.
The conditions are based on unitary operators commuting with the Hamiltonian and are associated with local symmetries that guarantee compact localized states and flatbands.
Beyond flatbands in lattice models, we can also apply our results to perturbed Hamiltonians where some internal states are not affected by additional perturbations (see Appendix~\ref{app:defects}).
We expect that the conditions derived here can be extended to design extraordinary states robust against local perturbations or environmental changes in a variety of coupled systems.

\section*{Acknowledgments}

The authors thank Emil Yuzbashyan for helpful discussions. 
We acknowledge financial support from the Institute for Basic Science in the Republic of Korea through the project IBS-R024-D1.

\appendix

\section{Uniform \(H_1\)}
\label{app:trivialh1}

The \(n \times n\) Hamiltonian for special cases with uniform \(H_1\) is
\begin{gather}
    \mh = H_0 + e^{-i k} H_1 + e^{i k} H_1^{\dagger},
\end{gather}
where \(H_0\) is the \(n \times n\) Hamiltonian with zero on-site energies and where all connected intra-cell hopping strengths are the same, i.e.,
\begin{gather}
    H_0 =
    \begin{pmatrix}
        \ddots & \vdots & \vdots & \vdots & \reflectbox{$\ddots$} \\
        \cdots & 0 & 1 & 1 & \cdots \\
        \cdots & 1 & 0 & 1 & \cdots \\
        \cdots & 1 & 1 & 0 & \cdots \\
        \reflectbox{$\ddots$} & \vdots & \vdots & \vdots & \ddots
    \end{pmatrix},
    H_1 =  
    \begin{pmatrix}
        \ddots & \vdots & \vdots & \vdots & \reflectbox{$\ddots$} \\
        \cdots & t & t & t & \cdots \\
        \cdots & t & t & t & \cdots \\
        \cdots & t & t & t & \cdots \\
        \reflectbox{$\ddots$} & \vdots & \vdots & \vdots & \ddots
    \end{pmatrix}.
\end{gather}

The \(2 \times 2\) case is the cross-stitch lattice in Section~\ref{subsec:examples1}.
Similar to the \(2 \times 2\) case, there is a trivial solution \(H_1=0\), which we dismiss.
For the nontrivial case, the Hamiltonian has \(n\) bands given by
\begin{align}
    E_\mathrm{D} &= (n-1) + 2 n t \cos{k}, \\
    \efb &= -1,
\end{align}
where \(E_\mathrm{D}\) and \(\efb\) represent dispersive and flat band energies, respectively. 
The number of dispersive and flat bands is \(1\) and \(n-1\), respectively. 
These special cases are also covered by our flatband conditions.

\section{Generic \(2 \times 2\) Hamiltonian: Deriving \(H_0\) and \(H_1\) from symmetries}
\label{app:generalizations1}

In the \(2 \times 2\) Hamiltonian case above, we only considered parity symmetric cases with two unitary operators, \(\sigma_0\) and \(\sigma_x\), commuting with \(H_0\). 
A generic \(2 \times 2\) Hermitian Hamiltonian can be written in terms of three Pauli matrices \(\sigma_x\), \(\sigma_y\), \(\sigma_z\) and the identity matrix \(\sigma_0\). 
Accordingly, the possible combinations of symmetries are as follows: 
\((\sigma_0, \sigma_x)\), \((\sigma_0, \sigma_y)\), \((\sigma_0, \sigma_z)\), \((\sigma_x, \sigma_y)\), \((\sigma_x, \sigma_z)\), \((\sigma_y, \sigma_z)\), 
\((\sigma_0, \sigma_x, \sigma_y)\), \((\sigma_0, \sigma_x, \sigma_z)\), \((\sigma_0, \sigma_y, \sigma_z)\), \((\sigma_x, \sigma_y, \sigma_z)\), 
and \((\sigma_0, \sigma_x, \sigma_y, \sigma_z)\). 
The \(2\times 2\) case we obtained in Section~\ref{subsec:examples1} corresponds to the first combination, \((\sigma_0, \sigma_x)\).
Below we consider all other possible combinations of symmetries.
The three combinations \((\sigma_0, \sigma_x)\), \((\sigma_0, \sigma_y)\), and \((\sigma_0, \sigma_z)\) give similar conditions\replaced{:}{}
\begin{gather}
    \left[\sigma_{0,x}, H_0\right] = 0,
    \left[\sigma_{0,x}, H_1\right] = 0
    ~~~~~ \mathrm{for} ~~ (\sigma_0, \sigma_x) 
    \label{eq:H0_sigmax}\\
    \left[\sigma_{0,y}, H_0\right] = 0,
    \left[\sigma_{0,y}, H_1\right] = 0
    ~~~~~ \mathrm{for} ~~ (\sigma_0, \sigma_y)
    \label{eq:H0_sigmay}\\
    \left[\sigma_{0,z}, H_0\right] = 0,
    \left[\sigma_{0,z}, H_1\right] = 0
    ~~~~~ \mathrm{for} ~~ (\sigma_0, \sigma_z)
    \label{eq:H0_sigmaz}
\end{gather}
from Eq.~(\ref{eq:comm}) and
\begin{gather}
    \sigma_0 H_1 = \pm \sigma_x H_1
    ~~~~~ \mathrm{for} ~~ (\sigma_0, \sigma_x)
    \label{eq:H1_sigmax}\\
    \sigma_0 H_1 = \pm \sigma_y H_1 
    ~~~~~ \mathrm{for} ~~ (\sigma_0, \sigma_y)
    \label{eq:H1_sigmay}\\
    \sigma_0 H_1 = \pm \sigma_z H_1
    ~~~~~ \mathrm{for} ~~ (\sigma_0, \sigma_z)
    \label{eq:H1_sigmaz}
\end{gather}
from Eq.~(\ref{eq:final_sol5}).
Parameterizing \(H_0\) and \(H_1\) as
\begin{gather}
    H_0 =
    \begin{pmatrix}
        \epsilon_{11} & \epsilon_{12} \\
        \epsilon_{21} & \epsilon_{22}
    \end{pmatrix}, 
    H_1 =  
    \begin{pmatrix}
        t_{11} & t_{12} \\
        t_{21} & t_{22}
    \end{pmatrix}
\end{gather}
and imposing the flatband conditions Eqs.~(\ref{eq:H0_sigmax})--(\ref{eq:H1_sigmaz}), we find the following solutions
\begin{align}
    \label{sigma_x}
    H_0 &=
    \begin{pmatrix}
        \epsilon_{11} & \epsilon_{12} \\
        \epsilon_{12} & \epsilon_{11}
    \end{pmatrix},
    H_1 =
    \begin{pmatrix}
        t_{11} & \pm t_{11} \\
        \pm t_{11} & t_{11}
    \end{pmatrix}, \\
    \label{sigma_y}
    H_0 &= 
    \begin{pmatrix}
        \epsilon_{11} & i\epsilon_{12} \\
        -i\epsilon_{12} & \epsilon_{11}
    \end{pmatrix},
    H_1 =
    \begin{pmatrix}
        t_{11} & \mp i t_{11} \\
        \pm i t_{11} & t_{11}
    \end{pmatrix}, \\
    H_0 &= 
    \begin{pmatrix}
        \epsilon_{11} & 0 \\
        0 & \epsilon_{22}
    \end{pmatrix},
    H_1 =
    \begin{pmatrix}
        t_{11} & 0 \\
        0 & 0
    \end{pmatrix}
    \mathrm{or} 
    \begin{pmatrix}
        0 & 0 \\
        0 & t_{22}
    \end{pmatrix}
    \label{sigma_z},
\end{align}
for the choices of symmetry operators \((\sigma_0, \sigma_x)\), \((\sigma_0, \sigma_y)\), and \((\sigma_0, \sigma_z)\), respectively.
The corresponding flatband energies are \(\epsilon_{11} \mp \epsilon_{12}\), \(\epsilon_{11} \mp \epsilon_{12}\), and \(\epsilon_{22}\) (or \(\epsilon_{11}\)), respectively.
Interestingly, the Hamiltonian of Eq.~\eqref{sigma_z} can be obtained by unitary transformations of the Hamiltonian of Eq.~\eqref{sigma_x}, i.e., detangling flatbands~\cite{Fla14}.

Considering the three combinations of \((\sigma_x, \sigma_y)\), \((\sigma_x, \sigma_z)\), and \((\sigma_y, \sigma_z)\) and using the same methods as above to resolve the hopping matrices,
we obtain \(H_0 = \epsilon_{11} \sigma_0\) and \(t_{ij} = 0~ (i, j = 1, 2)\) in all cases.
These are trivial cases of flatbands with no inter-cell hopping. 
As a consequence, the remaining combinations with three Pauli operators also give rise to trivial flatbands because these combinations always contain at least two Pauli matrices, similar to the case just discussed.
As a result, Eqs.~(\ref{sigma_x})--(\ref{sigma_z}) are \(2 \times 2\) Hermitian Hamiltonians with flatbands including trivial cases.

\begin{figure}
    \centering
    \includegraphics[width=1.0\linewidth]{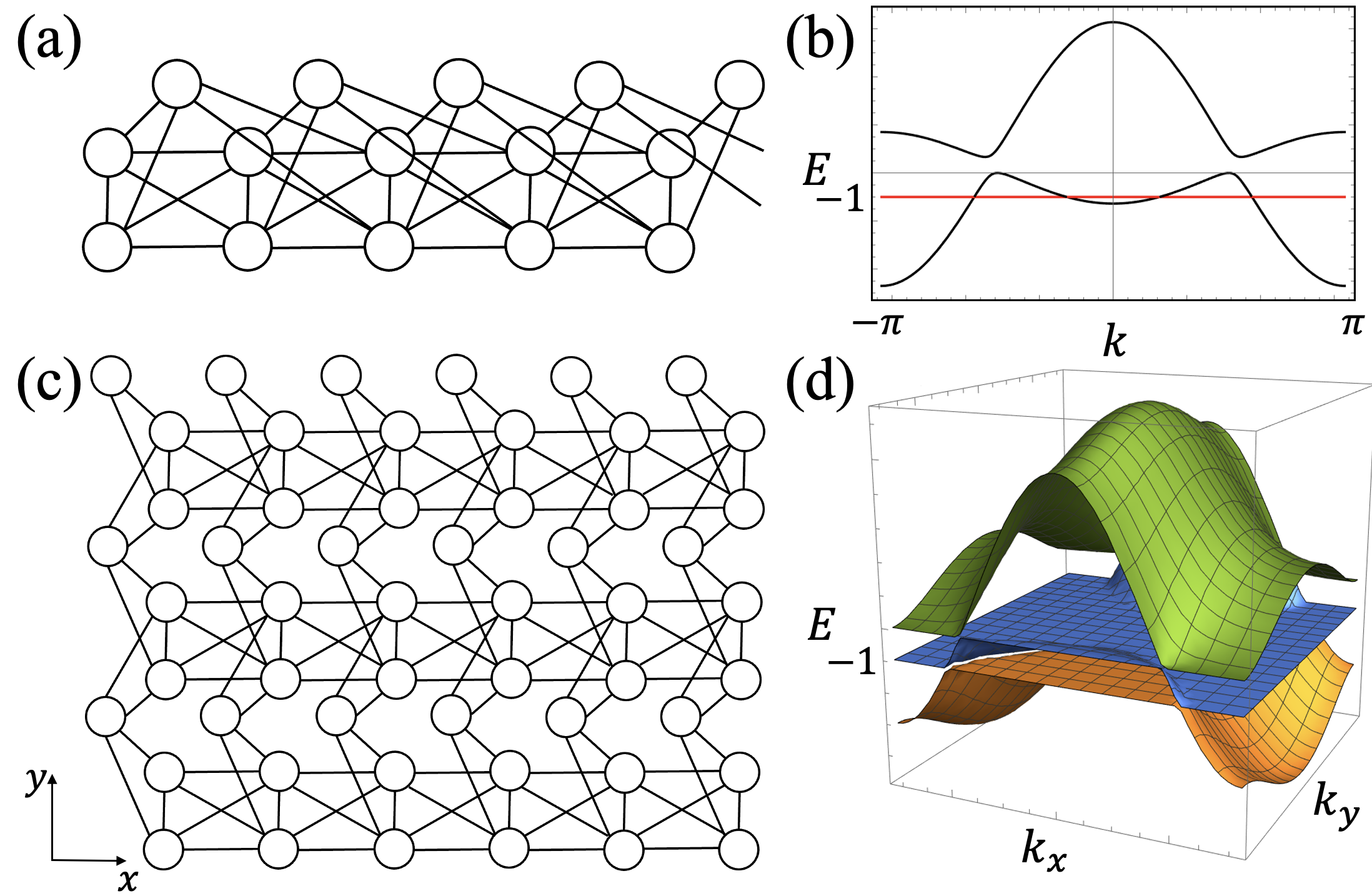}
    \caption{
        (a) 2D lattice model combining a cross-stitch lattice with a tunable diamond lattice. 
        (b) Energy bands of the 2D lattice model. 
        (c) 1D lattice model with different nearest and next-nearest hoppings. 
        (d) Energy bands of the lattice model of (c).
    }
    \label{figA1}
\end{figure}

\section{Hamiltonian with further neighbor hopping}
\label{app:ham_1d_lr}

In Section~\ref{subsec:generalizations3} we considered an example of a 1D flatband Hamiltonian with next-nearest hopping terms,
\begin{gather}
    H = H_0 + e^{-i k} H_1 + e^{i k} H_1^{\dagger} + e^{-2 i k} H_2 + e^{2 i k} H_2^{\dagger}.
    \label{App_LR_Ham}
\end{gather}
Imposing the symmetries $\sigma_0$ and $\sigma_1$ in Eq.~(\ref{eq:symm_long}) makes $H_0$, $H_1$, and $H_2$ take the forms as in Eqs.~(\ref{eq:h0_longrange}) and (\ref{eq:h1_longrange}). An example is a tunable diamond lattice with additional next-nearest hopping terms, which are described by the matrices
\begin{gather}
    H_0 =
    \begin{pmatrix}
        0 & 1 & 1 \\
        1 & 0 & 1 \\
        1 & 1 & 0
    \end{pmatrix},
    H_1 =
    \begin{pmatrix}
        0 & 0 & 0 \\
        0 & 1 & 1 \\
        0 & 1 & 1
    \end{pmatrix},
    H_2 =
    \begin{pmatrix}
        0 & 0 & 0 \\
        1 & 0 & 0 \\
        1 & 0 & 0
    \end{pmatrix}.
    \label{Long_Ham_H}
\end{gather}
Figure~\ref{figA1} (a) and (b) show this 1D lattice with different nearest and next-nearest hoppings and energy bands, among which one band is flat.

\section{Combined 2D Hamiltonians}
\label{app:combined}

In Section~\ref{subsec:generalizations2} we considered an example of a 2D flatband Hamiltonian,
\begin{gather}
    H = H_0 + e^{-i k_{x}} H_1 + e^{i k_{x}} H_1^{\dagger} + e^{-i k_{y}} H_2 + e^{i k_{y}} H_2^{\dagger}.
    \label{2D_Ham}
\end{gather}
Imposing symmetries $\sigma_0$ and $\sigma_1$ in Eq.~(\ref{eq:symm_2D}), $H_0$, $H_1$, and $H_2$ take the forms in Eqs.~(\ref{eq:h0_2d}), (\ref{eq:h1_2d}), and (\ref{eq:h1_2d_2}), respectively. As an example, a cross-stitch lattice with a tunable diamond lattice can be constructed from the same matrices as in Eq.~\eqref{Long_Ham_H}.
Figure~\ref{figA1} (c) and (d) show this 2D lattice model and energy bands, among which one band is flat.

\section{Hamiltonian with unperturbed internal states}
\label{app:defects}

Beyond flatbands in translationally lattice systems, our method can also be applied to a perturbed Hamiltonian,
\begin{gather}
    H = H_{0} + \delta H_{p},
\end{gather}
where \(H_{p}\) describes the perturbation and \(\delta\) is the perturbation strength.
If we obtain \(H_{p}\) satisfying our conditions from a given \(H_0\), some eigenstates of \(H\) are not affected by the perturbation as \(\delta\) increases.
For example, we consider the perturbed Hamiltonian with
\begin{align}
    H_0 =
    \begin{pmatrix}
        0 & 1 & 1 \\
        1 & 0 & 1 \\
        1 & 1 & 0
    \end{pmatrix},
    H_{p} =
    \begin{pmatrix}
        0 & 1 & 1 \\
        1 & 0 & 0 \\
        1 & 0 & 0
    \end{pmatrix}.
\end{align}
The eigenvalues are \(-1\) and \((1 \pm \sqrt{9+ 8 \delta (2+\delta)})/2\).
One of the eigenvalues is irrespective of the perturbation strength \(\delta\) since \(H_p\) satisfies the condition of Eq.~\eqref{eq:final_sol5}.

\bibliographystyle{apsrev4-2}
\bibliography{SP_flatbands_1}

\end{document}